\documentclass[aps,prl,twocolumn,superscriptaddress,groupedaddress]{revtex4}
\usepackage{graphicx,comment,float,braket,textcomp,amsmath,lipsum}  
\usepackage{dcolumn}   
\usepackage{bm}        
\usepackage{amssymb}   

\usepackage{color}

\begin{document}

\title{Probing Complex Decoherence Processes in Materials for Quantum Applications}

\author{Albert Liu}
\email{albert.liu@mpsd.mpg.de}
\affiliation{Brookhaven National Laboratory, Upton, New York, USA}

\author{Matthew W. Day}
\email{matthew.day@mpsd.mpg.de}
\affiliation{Max Planck Institute for the Structure and Dynamics of Matter, Hamburg, Germany}

\author{Steven T. Cundiff}
\email{cundiff@umich.edu}
\affiliation{Department of Physics, University of Michigan, Ann Arbor, Michigan, USA}


\vskip 0.25cm

\date{\today}

\begin{abstract}
    The primary consideration in developing new material platforms for quantum applications is to optimize coherence. Despite its importance, decoherence processes remains challenging to experimentally interrogate and quantify. In this Perspective, we first introduce the concept of decoherence in quantum systems and conventional techniques to assess decoherence at optical frequencies. We then introduce multidimensional coherent spectroscopy as a unique probe capable of revealing the full complexity of decoherence dynamics in realistic circumstances. To contextualize the techniques discussed here, demonstrative examples in two prototypical quantum systems, namely colloidal nanocrystals and vacancy centers in diamond, are provided.
\end{abstract}

\maketitle

Quantum technologies rely on the manipulation of two-level systems that serve as qubits, the quantum mechanical analogue of a classical bit. These systems may be embodied in a variety of material platforms such as superconducting Josephson junctions \cite{Kjaergaard2020}, semiconductor quantum dots \cite{Chatterjee2021}, vacancy centers in diamond \cite{Chatterjee2021,Wei2022,Chen2020}, and trapped atomic/ionic systems \cite{Wei2022,Kielpinski2002,Evered2023}. Regardless of the physical implementation, control over coupling between qubits and their dissipative environment is crucial to engineering systems for practical applications. For instance, thermal noise causes information loss in superconducting qubits \cite{Corcoles2011}, whereas coupling to background spins often causes loss of fidelity or coherence in vacancy center systems \cite{Barry2020,Williams2025}. 

Just as important is the ability to communicate between different qubits. While quantum information is most efficiently transmitted using optical photons \cite{Northup2014,Luo2023,Hu2023,Couteau2023}, practical qubits typically operate at microwave frequencies. The intermediate transduction of quantum information between these vastly different energy and time scales therefore introduces even more opportunities for dissipation to occur \cite{Lauk2020}. Photonic black-body radiation is also a major source of noise and error beyond single pairs of qubits in compute-heavy applications where a great number of qubits must be networked together for complex operations \cite{Corcoles2011}. Altogether, the interactions between qubits and their environment can destroy the quantum information contained in their fragile superpositions of states, a process known as {\it decoherence}.

Often, exponential loss of coherence is assumed in qubits interacting with a dissipative bath. However, this assumption is not necessarily true in many practical implementations of both individual and networks of qubits \cite{Li2019}. The possibility of non-exponential decoherence processes has proved particularly important in the context of quantum error correction as non-exponential (non-Markovian, as will be detailed later) decoherence processes retain memory of a qubit's prior states, possibly allowing for correction of the qubit's present state in question \cite{Champagne2020}. Dynamical decoupling schemes \cite{Ezzell2023} that have proven crucial to quantum state preservation also depend critically on non-Markovianity \cite{Addis2015}. Yet to properly take advantage of such complex processes, the details of system-bath coupling must be understood. 


What is urgently needed, then, is the widespread ability to probe complex decoherence processes for rational engineering of quantum technologies and their material platforms. While such efforts are fairly mature at microwave frequencies \cite{Bylander2011}, comparatively sparse effort has been spent on developing probes of dissipation in optical frequency systems. 

This perspective discusses optical spectroscopies as a way to understand the system-bath interactions in candidate high-energy qubit systems, with a focus on solid state materials due to their more complex decoherence processes. We begin with an introduction to decoherence as a general concept, then followed by a discussion of conventional spectroscopies that are often employed to characterize decoherence. We finish our perspective by detailing the advantages of multidimensional coherent spectroscopy in characterizing decoherence, and more generally the underlying system-bath coupling in two different classes of candidate quantum systems: colloidal quantum dots and vacancy centers in diamond.

\section{Introduction to Decoherence in Quantum Systems}

We begin with a brief overview of decoherence from the perspective of atomic physics, from which the concept arose. The simplest model of an atom is that of a two-level system consisting of a `ground' energy level $\Ket{0}$ and `excited' energy level $\Ket{1}$ with energies $E_0$ and $E_1$ respectively, corresponding to the Hamiltonian
\begin{align}\label{Hamiltonian}
    \hat{H} = E_0\Ket{0}\Bra{0} + E_1\Ket{1}\Bra{1}.
\end{align}
An arbitrary quantum state of the atom $\Ket{\Psi(t)}$ will evolve in time
\begin{align} \label{eq_AtomicState}
    \Ket{\Psi(t)} = a_0e^{-\frac{i}{\hbar}E_0t}\Ket{0} + a_1e^{-\frac{i}{\hbar}E_1t}\Ket{1},
\end{align}
where $a_0$ and $a_1$ are complex coefficients that define the initial state at $t = 0$. In many cases, one is interested in the atomic dipoles that lead to light absorption/emission. For the general state defined in (\ref{eq_AtomicState}), the time-dependent expectation value of the dipole operator $\hat{\mu}$ is
\begin{align}
    \Braket{\Psi(t)|\hat{\mu}|\Psi(t)} &= a_0^*a_1\mu_{01}e^{-i\omega_{10}t} + a_0a_1^*\mu_{10}e^{i\omega_{10}t},
\end{align}
which scales with a transition dipole moment $\mu_{ij} = \Braket{i|\hat{\mu}|j}$ and rotate at a frequency $\omega_{10} = E_{10}/\hbar$, where $E_{10} = E_1 - E_0$. These rotating terms are only present if both $a_0$ and $a_1$ are non-zero, when $\Ket{\Psi(t)}$ is a {\it quantum superposition} of states, and are therefore the most straightforward manifestation of \textit{quantum coherence}.

To examine quantum coherence more closely, it is useful to recast the system dynamics in terms of the density matrix
\begin{align}\label{DensityMatrixDefinition}
    \nonumber \rho(t) &= \Ket{\Psi(t)}\Bra{\Psi(t)}\\
    \nonumber &= |a_0|^2\Ket{0}\Bra{0} + |a_1|^2\Ket{1}\Bra{1}\\ 
    \nonumber &\hspace{1cm}+ a_1a_0^*e^{-i\omega_{10}t}\Ket{1}\Bra{0} + a_0a_1^*e^{i\omega_{10}t}\Ket{0}\Bra{1}\\
    &= \rho_{00}\Ket{0}\Bra{0} + \rho_{11}\Ket{1}\Bra{1} + \rho_{10}\Ket{1}\Bra{0} + \rho_{01}\Ket{0}\Bra{1},
\end{align}
where we can now identify the off-diagonal density matrix elements $\rho_{ij}(t) = a_ia_j^*e^{-i\omega_{ij}t}$ as direct representations of quantum coherence. Here we have neglected interactions of this two-level system with any other degrees of freedom that induce relaxation of the density matrix, for example with other level systems, vacuum electromagnetic fields (responsible for spontaneous emission \cite{Sanchez-Mondragon1983}), or a generic thermal bath. The above time-dependence of $\rho_{ij}(t)$ is, however, derived under the assumption of a static two-level system in which $\omega_{10}$ does not vary. If $\omega_{10}$ changes in time, for example through interacting with a surrounding environment, the time-dependence of off-diagonal elements $\rho_{ij}(t)$ may be simply obtained from the von Neumann equation of motion for a Hamiltonian analogous to that of (\ref{Hamiltonian}) with time-dependent eigenenergies $E_i(t) = \braket{i|\hat{H}(t)|i}$:
\begin{align}
    \nonumber \frac{d}{dt}\rho_{ij}(t) &= -\frac{i}{\hbar}\Braket{i|\left[\hat{H}(t),\rho(t)\right]|j}\\
    &= -i\omega_{ij}(t)\rho_{ij}(t),
\end{align}
where the resonance frequency $\omega_{ij}(t)$ is now also time-dependent. We find the formal solution by integration:
\begin{align}\label{CrossDiagonalTimeDependence}
    \nonumber \rho_{ij}(t) &= \rho_{ij}(t = 0)e^{-i\int^t_0\omega_{ij}(\tau)d\tau}\\
    &= \rho_{ij}(t = 0)e^{-i\overline{\omega_{ij}}t}e^{-i\int^t_0\delta\omega_{ij}(\tau)d\tau}
\end{align}
in which we've explicitly separated $\omega_{ij}(t) = \overline{\omega_{ij}} + \delta\omega_{ij}(t)$ into its time-averaged value and its fluctuations. For static $\omega_{ij}(t) = \omega_{ij}$, the above solution reduces to that previously obtained in (\ref{DensityMatrixDefinition}). However, to see the effect of fluctuations in a real systems, we take an ensemble average over the environmental degrees of freedom
\begin{align}
        \nonumber \Braket{\rho_{ij}(t)} &= \rho_{ij}(t = 0)e^{-i\overline{\omega_{ij}}t}\Braket{e^{-i\int^t_0\delta\omega_{ij}(\tau)d\tau}}\\
        &\approx \rho_{ij}(t = 0)e^{-i\overline{\omega_{ij}}t}e^{-\int^t_0\int^\tau_0\Braket{\delta\omega_{ij}(\tau')\delta\omega_{ij}(0)}d\tau' d\tau}
\end{align}
where $\Braket{\dots}$ indicates the ensemble average and the final expression written in terms of a fluctuation correlation function is derived via cumulant expansion and truncating at second-order \cite{Mukamel1999_Book}. We now see that fluctuations $\delta\omega_{ij}$ cause deviations from pure oscillatory behavior, and generally cause $\rho_{ij}(t)$ to decrease over long timescales (discussed in more detail below). This loss in quantum coherence is known as {\it decoherence}, and understanding the underlying mechanisms of decoherence is central to engineering materials for quantum applications.

\section{Conventional Methods of Measuring Decoherence}

Conventional methods of measuring decoherence may be categorized as either {\it linear} or {\it nonlinear} spectroscopies. Quantum coherence may manifest at frequencies across the electromagnetic spectrum, but here we restrict our discussion to techniques that measure quantum coherence at optical frequencies.

\begin{figure}[b]
    \centering
    \includegraphics[width=0.3\textwidth]{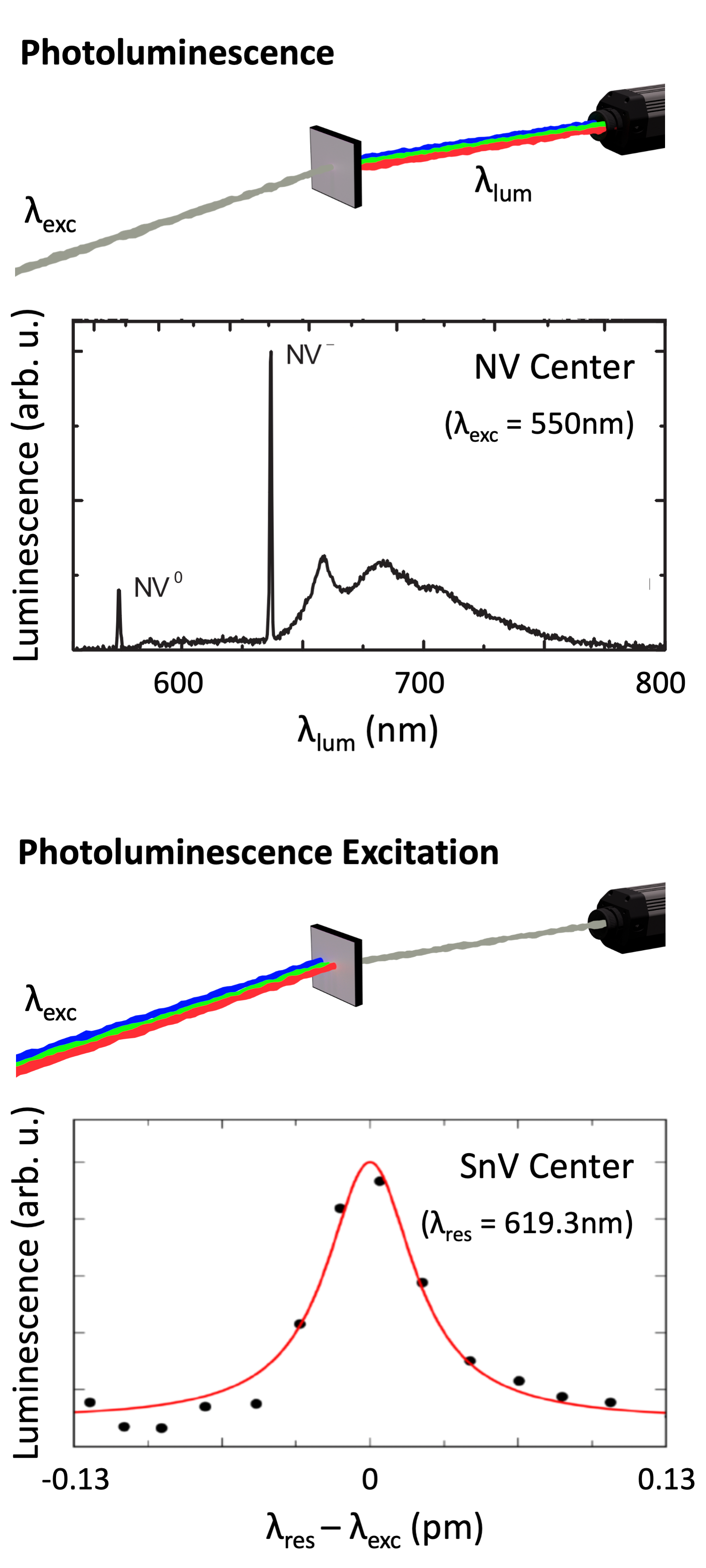}
    \caption{(Top) Schematic photoluminescence spectroscopy and measurements of a single nitrogen-vacancy (NV) center in diamond at 10 K. Narrow zero-phonon lines are observed from two different charge states NV$^0$ and NV$^-$. Figure adapted from \cite{Beha2012}. (Bottom) Schematic photoluminescence excitation spectroscopy and measurements of a single tin-vacancy (SnV) center in diamond at 1.7 K, which isolated luminescence into the SnV phonon sideband. Figure adapted from \cite{Rugar2020}.}
    \label{Fig1}
\end{figure}

\subsection{Linear Spectroscopies}

Photoluminescence spectroscopy (PL), a subset of luminescence/fluorescence spectroscopies, involves photoexcitation at photon energies at or above a material resonance which leads to subsequent light emission. The linewidth of a peak in the resultant emission spectrum can be related to an effective dephasing time $T_2$ of the underlying transition \cite{Gammon1996}. The PL spectrum of a single nitrogen-vacancy (NV) center in diamond for excitation wavelength $\lambda_{exc}$ and as a function of luminescence wavelength $\lambda_{lum}$ is shown in the top panel of Figure \ref{Fig1}, in which two peaks corresponding to the neutral state (NV$^0$) and negatively-charged state (NV$^-$) optical transitions are observed \cite{Beha2012}. PL measurements from single emitters have also been performed in quantum dots \cite{Brunner1992,Gammon1996}, nanocrystals \cite{Yin2017}, and molecules \cite{Moerner2015,Kondo2017}.

For isolated two-level systems, absorption and luminescence spectra provide identical information about decoherence \cite{Mukamel1999_Book}. While PL from single emitters is now routinely accomplished, directly measuring their weak absorption spectra is still prohibitively challenging. Photoluminescence excitation spectroscopy (PLE) provides an indirect method for characterizing absorption spectra of single emitters, in which the luminescence strength is measured as a function of the wavelength of a narrowband excitation light source. The PLE spectrum of a single tin-vacancy (SnV) center in diamond is shown in the bottom panel of Figure \ref{Fig1} as a function of detuning from the resonance wavelength $\lambda_{res} = 619.3$ nm, in which an optical transition was resolved with a linewidth in the MHz frequency range. PLE has also been used to measure linewidths in other systems such as alternate vacancy centers in diamond \cite{Chu2014,Häußler2017}, colloidal nanocrystals \cite{Norris1996,Htoon2004,Tonti2004}, and vacancy centers in HBN \cite{Hoese2020}.

However, the above techniques are limited when the optical properties of an emitter vary in time, known as spectral diffusion, and/or when measuring decoherence in ensembles of emitters. In ensembles, static disorder in the resonance frequency, known as {\it inhomogeneous broadening}, obscures the true linewidth due to decoherence \cite{Empedocles1996}. This extrinsic line-broadening is typically impenetrable to linear spectroscopies, with the exception of fluorescence line narrowing (FLN). FLN is a variant of PL that involves narrowband excitation of a single frequency group of emitters, resulting in a luminescence spectrum that is far more homogeneous \cite{Nirmal1994}. However FLN is susceptible to secondary photoexcitation by the initial luminescence, restricting measurements to the long-wavelength (low-energy) portion of the inhomogeneous distribution. To circumvent inhomogeneous broadening entirely, nonlinear spectroscopies are needed.

\subsection{Nonlinear Spectroscopies}

\begin{figure}[t]
    \centering
    \includegraphics[width=0.3\textwidth]{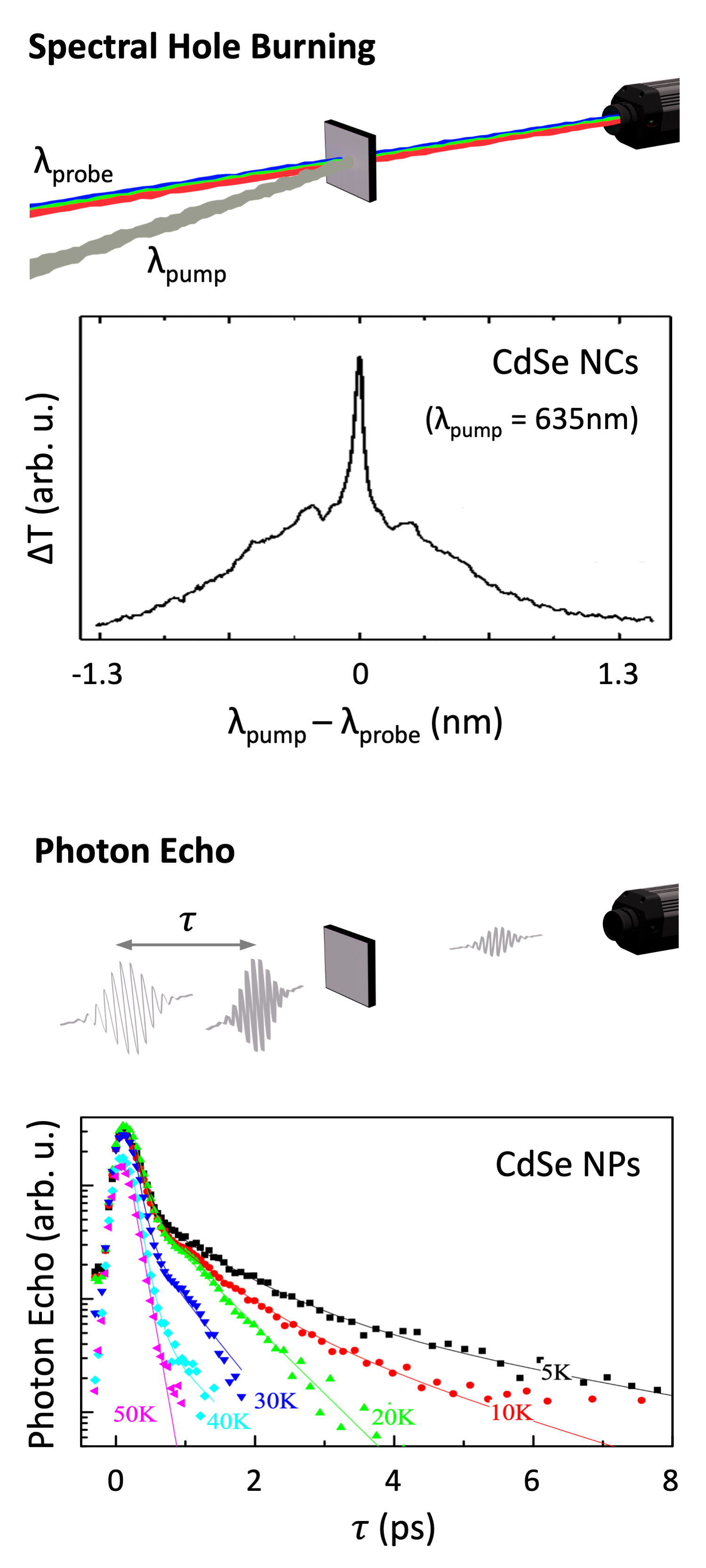}
    \caption{(Top) Schematic spectral hole burning and measurements of CdSe colloidal nanocrystals at 10 K. The pump field is fixed at 635 nm, while the probe field is scanned from shorter to longer wavelength to reveal the homogeneous lineshape. Figure adapted from \cite{Palinginis2003}. (Bottom) Schematic photon echo spectroscopy and measurements of CdSe colloidal nanoplatelets at the temperatures indicated. The photon echo field amplitude is plotted. Figure adapted from \cite{Naeem2015}.}
    \label{Fig2}
\end{figure}

Spectral hole burning (SHB) \cite{Moerner1998_Book} is a technique that, in the context of characterizing decoherence, is conceptually similar to FLN. In both techniques, a narrowband 'pump' of wavelength $\lambda_{pump}$ is used to isolate a specific frequency group of emitters from an inhomogeneously broadened distribution. Rather than subsequent luminescence measured in FLN, SHB involves measuring pump-induced changes in the sample absorption spectrum with a 'probe' field of wavelength $\lambda_{probe}$ (typically from another tunable narrowband laser). The SHB spectrum of an ensemble of CdSe nanocrystals is shown in the top panel of Figure~\ref{Fig2}, in which the zero-phonon line of an exciton transition is observed superimposed on a broader lineshape due to exciton-phonon coupling.

Perhaps a more natural way of measuring the temporal phenomenon of decoherence is in the time-domain. Photon echo spectroscopy (PES) \cite{Yajima1979} uses two (or three) pulses of light, separated by a time delay $\tau$, to generate a photon echo signal (an optical analogue of spin echo signals in nuclear magnetic resonance) that decays with increasing inter-pulse time delay, with a decay rate equal to the effective dephasing time $1/T_2$. The generation process of photon echoes may be intuitively understood as the second (and third) pulse initiating a time-reversal operation that `rephases' the resonance frequency disorder at a time $t = \tau$. For further details, we refer the reader to more thorough treatments of photon echoes on the Bloch sphere \cite{AllenEberly_Book,Keeler2010_Book}. Two-pulse photon echo measurements of CdSe nanoplatelets are shown in the bottom panel of Figure~\ref{Fig2}, in which decoherence is directly measured in the time-domain at femto- to picosecond timescales.

\section{Multidimensional Coherent Spectroscopy as a Probe of Complex Decoherence}

In the previous section, conventional spectroscopic techniques that measure effective dephasing rates of both single emitters and ensembles were described. However, these techniques often obscure the full complexity of decoherence in many quantum-relevant systems. For example, invoking an effective dephasing time $T_2$ implicitly assumes exponential decoherence (the so-called {\it Markovian limit}), which is usually violated in systems with strong system-bath coupling at low temperatures. Characterizing decoherence becomes further complicated in systems with dynamic environments, resulting in correspondingly dynamic Hamiltonians that affect decoherence.

Here, we introduce multidimensional coherent spectroscopy (MDCS) as a probe of complex decoherence. MDCS \cite{HammZanni2012_Book,Cundiff2023_Book} is a generalization of four-wave mixing spectroscopies (such as spectral hole burning and photon echo spectroscopy detailed above) that is capable of resolving a nonlinear optical response along its multiple frequency dimensions simultaneously. An experimental schematic of MDCS is shown in Figure~\ref{Fig3}, in which three excitation pulses with variable inter-pulse time delays $\tau$ and $T$ generate a four-wave mixing signal that emits along the real laboratory time $t$. Simultaneous Fourier transform of the signal along a pair of time variables then results in a two-dimensional spectrum that yields different physics depending on the Fourier axes chosen. Below, we introduce several unique capabilities of MDCS with demonstrative examples from our own research.

\subsection{Disentangling Decoherence and Extrinsic Dephasing}

The primary advantage of the nonlinear spectroscopies detailed above is in measuring decoherence in the presence of inhomogeneous broadening. Both SHB and PES provide incomplete information concerning decoherence of an ensemble however, returning the dephasing time of a single frequency group and the average of the entire frequency distribution (in the form of non-exponential decay dynamics) respectively. MDCS overcomes the limitations of these techniques in spectra obtained by Fourier transform along the time variables $\{\tau,t\}$ (for the photon echo phase-matching condition). The signal evolutions along $\tau$ and $t$ reflect absorption and emission dynamics respectively, reflected in the names of their transform axes, and the power of MDCS may be intuitively understood from correlating the absorption and emission dynamics.

\begin{figure*}
    \centering
    \includegraphics[width=1\textwidth]{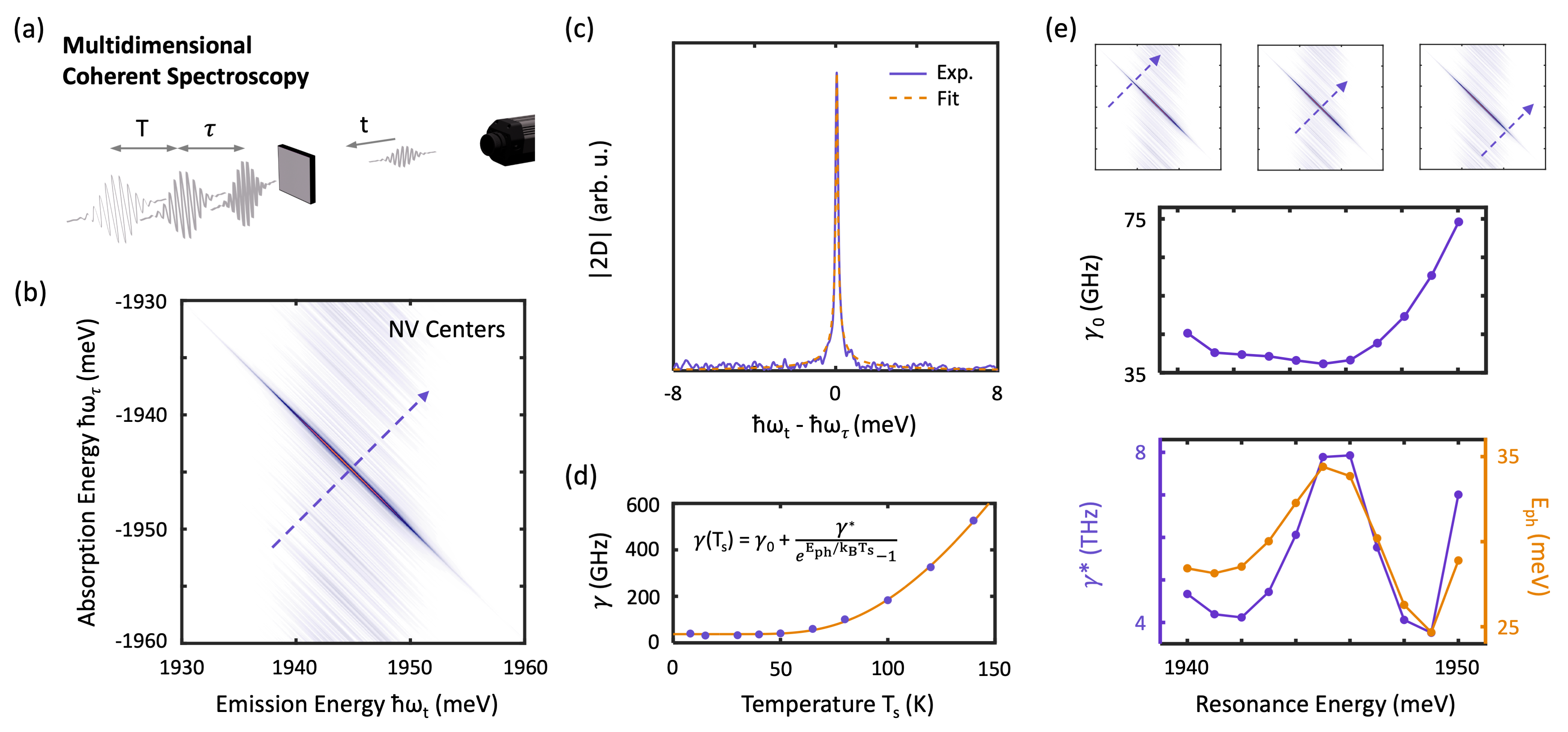}
    \caption{(a) Schematic of MDCS, in which three excitation pulses of variable inter-pulse time delays $\tau$ and $T$ generate an emitted four-wave mixing signal as a function of evolving real time $t$. (b) Single-quantum MDCS spectrum of the NV$^-$ zero-phonon line from an NV center ensemble at 15 K. (c) Slice of the single-quantum spectrum in (b) along the arrow indicated. (d) Dependence of the dephasing rate $\gamma$ on sample temperature $T_s$ extracted from slice fits, revealing broadening due to exciton-phonon coupling. (e) Resonance energy dependences of the thermal broadening parameters defined in the inset of (d). Figure adapted from \cite{Liu2021_MQT}.}
    \label{Fig3}
\end{figure*}

One such 'single-quantum' (referring to $\hbar\omega_\tau$ equal to a single unit of excitation photon energy) spectrum of a nitrogen-vacancy (NV) center ensemble in diamond is plotted in Figure~\ref{Fig3}b, which exhibits a single peak elongated in the $|\hbar\omega_\tau| = |\hbar\omega_t|$ direction due to dominant inhomogeneous broadening. The lineshape along the perpendicular direction indicated by an arrow reflects the homogeneous lineshape of a given frequency group (plotted in Figure~\ref{Fig3}c for a resonance energy of 1945 meV). Lineshape fits readily provide the associated dephasing rate $\gamma = 1/T_2$ (one to two orders of magnitude smaller than the inhomogeneous dephasing rate \cite{Liu2021_MQT}), which is then plotted as a function of temperature in Figure~\ref{Fig3}d. The observed trend fits well to a Boltzmann activation behavior due to exciton-phonon coupling with fit parameters $\gamma_0$ (zero-temperature dephasing rate), $\gamma^*$ (exciton-phonon coupling strength), and $E_{ph}$ (phonon energy). The ability of MDCS to resolve all frequency groups simultaneously further allows characterization of all three of these parameters as a function of resonance energy, which is plotted in Figure~\ref{Fig3}e. In contrast, this variation in dephasing rate would simply result in non-exponential decay in PES measurements, which obscures even the intrinsic dephasing rate. We note that this information is in principle obtainable via SHB, but would require tedious two-dimensional scanning of both pump and probe wavelengths \cite{Purchase2009}.

\subsection{Ultrafast Spectral Diffusion}

Decoherence is usually thought of as a quantum superposition of states interacting with the given spectral function of its environment fluctuations. Yet the environment itself may be dynamical and necessitate a time-dependent Hamiltonian. This phenomenon is routinely observed in nanocrystals and vacancy centers as resonance frequency fluctuations, known as {\it spectral diffusion}, on slow (ms - s) timescales \cite{Neuhauser2000,Fu2009,Plakhotnik2010,Chu2014}. For engineering materials for quantum applications however, probing spectral diffusion on ultrafast timescales relevant to quantum coherence is of primary importance.

\begin{figure}[b]
    \centering
    \includegraphics[width=0.4\textwidth]{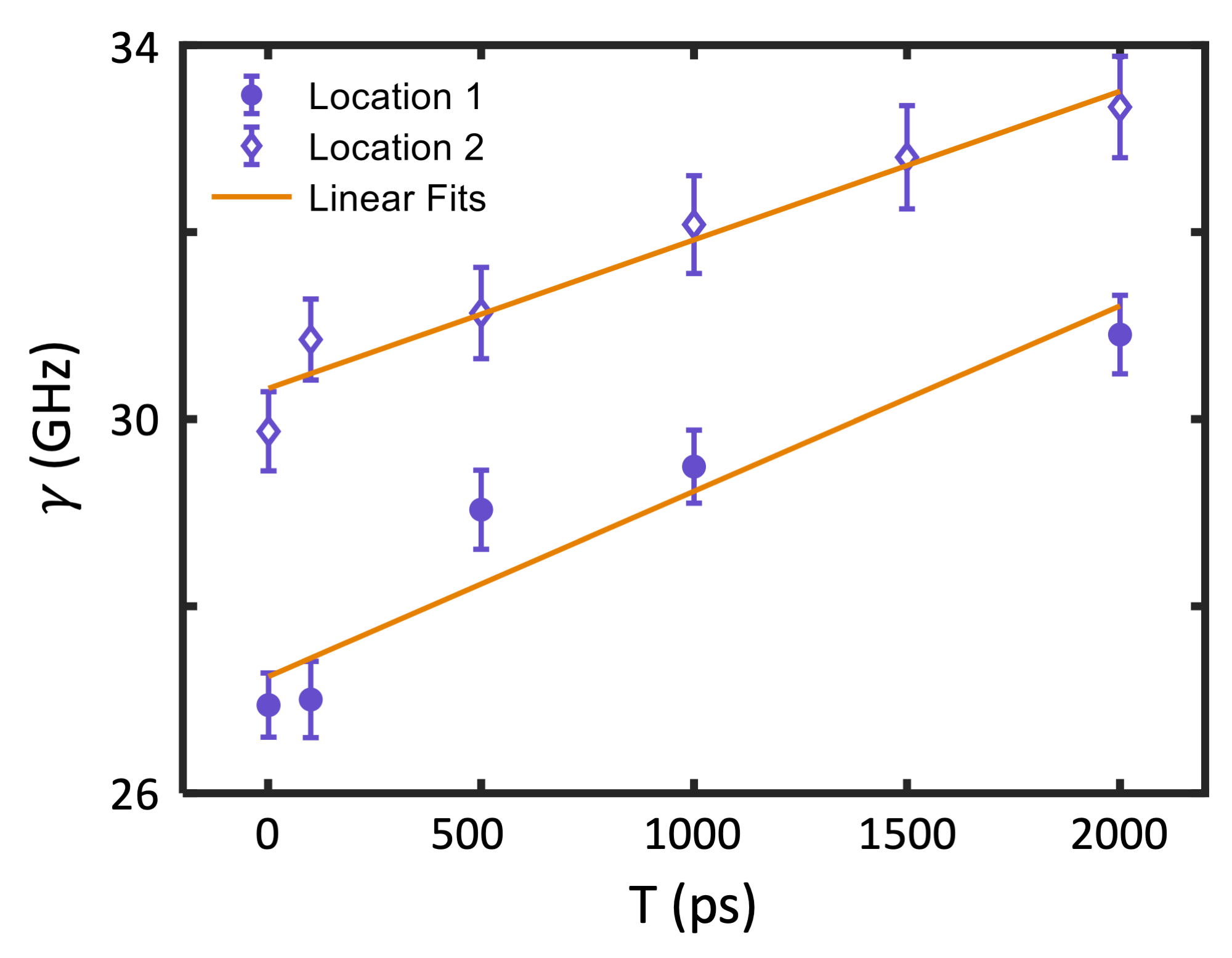}
    \caption{Measurements of an equivalent dephasing rate $\gamma$ in the same NV center sample studied in Figure~\ref{Fig3} at 10 K as a function of the time delay $T$. The monotonic increase reveals ultrafast spectral diffusion, and measurements at two locations are fit to slopes of 1.98 MHz/ps and 1.59 MHz/ps. Figure adapted from \cite{Liu2021_MQT}.}
    \label{Fig4}
\end{figure}

Single-quantum MDCS spectra provide direct access to ultrafast spectral diffusion by correlating absorption and emission dynamics. In the presence of incoherent resonance frequency fluctuations, absorption and emission frequency become increasingly uncorrelated along the waiting time delay $T$ (during which population dynamics evolve) which manifests as effective line-broadening. Ultrafast spectral diffusion measurements of NV centers are presented in Figure~\ref{Fig4}, in which the effective dephasing rate $\gamma$ corresponding to the homogeneous linewidth is observed to monotonically increase with $T$. Two different locations on a single sample both reveal spectral diffusion rates in the MHz/ps range.

Similar physics has been studied by MDCS in a variety of material systems ranging from nanocrystals \cite{Seiler2019,Nguyen2023} to quantum wells \cite{Singh2016,Singh2017} and quantum dots \cite{Wigger2020}. We also note the development of photon correlation spectroscopies \cite{Sallen2010,Lubin2022} that can also access spectral diffusion on ultrafast timescales, albeit only in single emitters \cite{Wolters2013}.

\begin{figure}[H]
    \centering
    \includegraphics[width=0.35\textwidth]{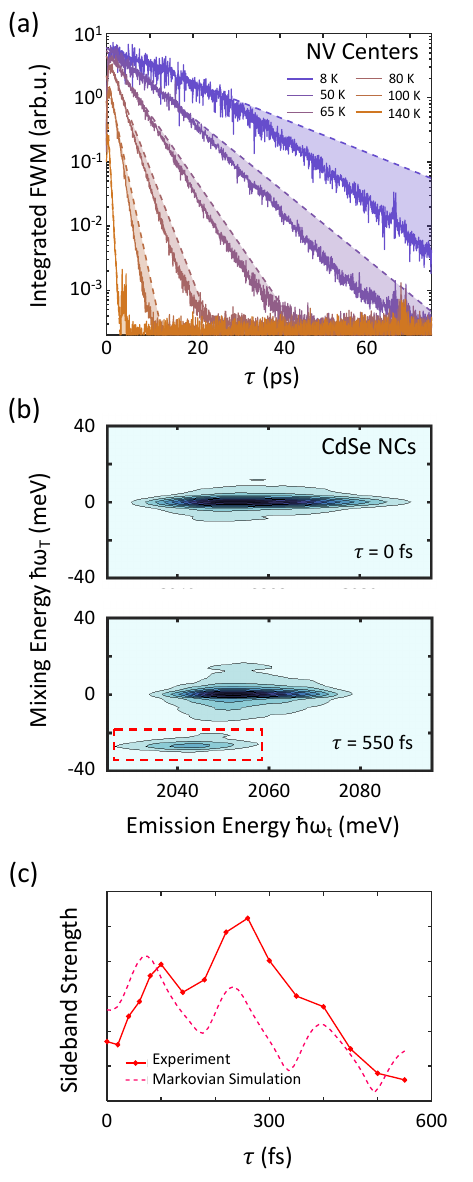}
    \caption{(a) Photon echo spectroscopy measurements of the same NV center sample studied in Figures~\ref{Fig3} and \ref{Fig4} as a function of temperature. Dashed lines represent an exponential fit to the early-time dynamics, and the deviation from experimental measurements (shaded areas) indicate non-Markovian dephasing at low temperatures. (b) Zero-quantum MDCS spectra of CdSe nanocrystals at 20 K at two time delays $\tau$. A sideband appears at large $\tau$ at a mixing energy $\hbar\omega_T \approx 26$ meV, the LO phonon energy of CdSe. (c) The sideband intensity integrated over the red box in (b) reveals non-Markovian dephasing dynamics of vibrationally-coupled interband coherences. Simulated Markovian dephasing dynamics are shown for comparison. Figures adapted from \cite{Liu2021_MQT} and \cite{Liu2019_PRL}.}
    \label{Fig5}
\end{figure}

\subsection{Non-Markovian Decoherence}

The discussions so far have considered situations in which decoherence is exponential and spectral lineshapes are Lorentzian. Recalling the microscopic origin of decoherence (ignoring $T_1$ relaxation) as transition frequency fluctuations, we can make the assumption that the fluctuations only exhibit instantaneous correlation, corresponding to the frequency correlation function $\Braket{\delta\omega_{ij}(t)\delta\omega_{ij}(0)} = \gamma\delta(t)$. In this case the time-dependence of $\rho_{ij}(t)$, the off-diagonal density matrix elements, is then found from (\ref{CrossDiagonalTimeDependence}) to be
\begin{align}\label{MarkovianCorrelationFunction}
    \nonumber \Braket{\rho_{ij}(t)} = \rho_{ij}(t = 0)e^{-i\overline{\omega_{ij}}t}e^{-\gamma t}.
\end{align}
We therefore identify a direct correspondence between a vanishing correlation time (Markov approximation) of the transition frequency fluctuations and exponential decoherence at a rate $\gamma$. For this reason, exponential decoherence is sometimes referred to as {\it Markovian decoherence}.

At elevated temperatures decoherence usually occurs in the Markovian limit described above. At cryogenic temperatures however, the regime at which most quantum technologies operate, non-Markovian decoherence can occur. Intuitively, non-Markovian decoherence can be understood as information backflow from a system to its `bath', which introduces memory into the bath-induced transition energy fluctuations. In recent years the idea of exploiting non-Markovianity for quantum applications has attracted increasing attention, for example in enhancing quantum control \cite{Reich2015}, quantum memories \cite{Man2015}, and preserving quantum coherence \cite{Miller2022}. Yet the difficulty in experimentally characterizing non-Markovianity continues to be an obstacle to harnessing such effects.

As the experimental signatures of non-Markovian decoherence are usually quite subtle, and easily obscured by disorder, it is best-studied in the time-domain with photon echoes. Exemplary PES measurements on NV centers are shown in Figure~\ref{Fig5}a, in which deviations from exponential behavior are observed at low temperatures. Besides direct measurement of the photon echo intensity, photon echo peak-shift measurements \cite{Joo1996,Oh2011} have also been used to measure the system-bath correlation function and signatures of non-Markovianity therein \cite{Lorenz2008}. However, we reiterate that photon echo techniques resolve the ensemble-averaged homogeneous response, which becomes confounding for baths with significant structure or for level systems involving more than a single optical transition. This limitation calls for alternative methods to study non-Markovianity in more complex systems.

As a Fourier transform spectroscopy MDCS resolves decoherence in both the time and frequency domains, offering the unique capability to extract dynamics of specific frequency groups. This concept was demonstrated in 'zero-quantum' spectra of CdSe nanocrystals, obtained by Fourier transform along the time variables $\{T,t\}$, plotted in Figure~\ref{Fig5}b. By isolating vibrational coherence dynamics resulting from exciton coupling to longitudinal-optical phonons of the CdSe lattice, non-Markovian decoherence of excitons in a vibrationally-excited state were observed along the time delay $\tau$ as shown in Figure~\ref{Fig5}c. While MDCS was used here to reveal signatures of non-Markovian decoherence in the time-domain, it is also useful for studying non-Markovianity in the frequency-domain as well. We expand on these frequency-domain signatures in the following section.

\subsection{Characterizing the Spectral Density of System-Bath Coupling}

In addition to its potential applications in quantum technologies, non-Markovian decoherence is also useful as a window into the microscopic mechanisms of system-bath coupling. In the Markovian limit all decoherence mechanisms contribute to an effective dephasing rate $\gamma = 1/T_2$, rendering exponential dephasing agnostic to even drastically differing models of system-bath interactions. On the other hand, non-Markovian decoherence reveals the underlying structure of system-bath coupling through their unique spectral lineshapes.

\begin{figure}[t]
    \centering
    \includegraphics[width=0.35\textwidth]{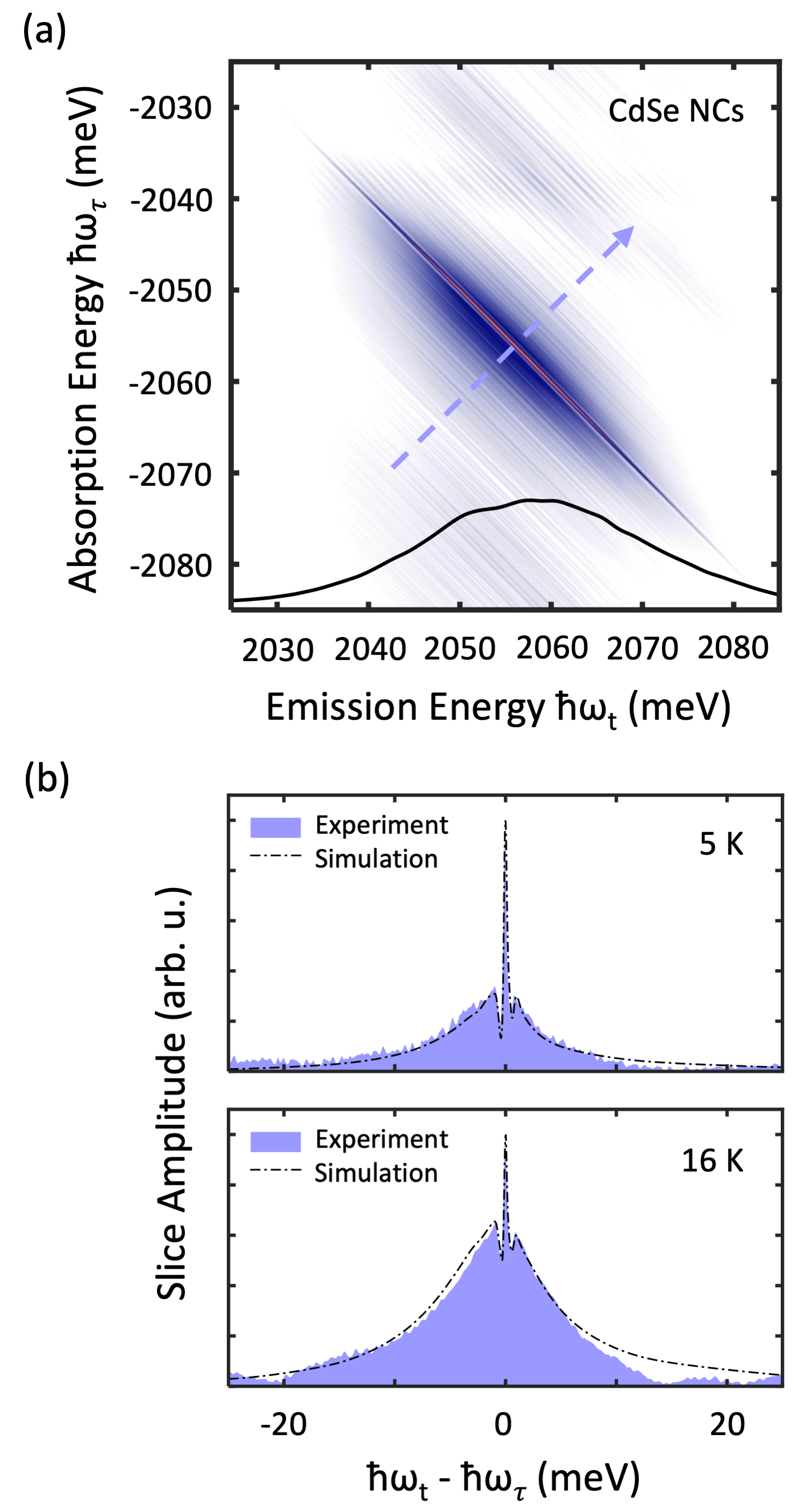}
    \caption{(a) Single-quantum MDCS spectra of CdSe nanocrystals at 5 K. The optical response is windowed by the excitation laser spectrum, shown by the black curve. (b) Comparison of slices taken of experimental and simulated single-quantum spectra along the arrow indicated in (a) at two temperatures. Spectral windowing of the simulated single-quantum spectra was incorporated using the respective experimental excitation laser spectra. Figure adapted from \cite{Liu2019_JPCL}.}
    \label{Fig6}
\end{figure}

Returning to the system of CdSe nanocrystals from the previous section, in which non-Markovian dephasing induced by discrete optical modes was considered, we now examine the exciton homogeneous lineshape isolated in single-quantum spectra (plotted in Figure~\ref{Fig6}a). By taking a slice along the direction indicated by the arrow in Figure~\ref{Fig6}a, a strongly non-Lorentzian homogeneous lineshape is observed that is attributed to exciton coupling to the CdSe acoustic phonon bath. The lineshapes are examined more closely in the slices at two temperatures plotted in Figure~\ref{Fig6}b, which exhibit two notable features. A sharp 'zero-phonon line' with a Lorentzian lineshape is superimposed on a broad asymmetric pedestal which varies dramatically with temperature. 

The lineshape of this pedestal is strongly non-Lorentzian, indicating non-Markovian decoherence \cite{Liu2024_AQT}, and may be directly related to the {\it spectral density} of the exciton-phonon coupling \cite{Nitzan2006_Book}
\begin{align}
    J(\omega) = \frac{\pi D(\omega)c^2(\omega)}{2\omega},
\end{align}
proportional to the phonon density of states $D(\omega)$ and the frequency-dependent exciton-phonon coupling strength $c^2(\omega)$. From its direct relation to the transition energy correlation function \cite{Mukamel1999_Book}, the spectral density may be loosely interpreted as the frequency spectrum of bath-induced transition energy fluctuations. Simulation of single-quantum spectra with the spectral density derived for a spherical nanocrystal \cite{Luker2017} are compared to experiment in Figure~\ref{Fig6}b, which demonstrates the powerful ability of MDCS to characterize system-bath coupling \cite{Liu_2023_ES} in materials for quantum applications.

\subsection{Interaction-Induced Decoherence}

Our discussion so far has considered non-interacting two-level systems coupled to a thermal bath. Yet more generally, different members of an ensemble may also mutually influence the coherence loss of their individual quantum states. From the perspective of quantum applications, it is crucial to distinguish this interaction-induced decoherence from other sources of dephasing. 
For example, dephasing due to ensemble disorder is not a resource for useful information exchange or manipulation. However, interacting emitters can be jointly controlled or entangled \cite{Unold2005,Deleseluc2017}, providing a potential resource for quantum information processing \cite{Brennen2000}. Below we demonstrate how MDCS can be used to identify subtle interactions between quantum systems in an ensemble of negatively-charged silicon vacancy centers (SiV$^-$) in diamond.

In the limit of a single SiV$^-$ emitter, the optical transitions possess coherence times exceeding several nanoseconds \cite{Becker2017,Rogers2014}. Yet when placed among an ensemble, luminescence measurements indicate that SiV$^-$ coherence times shorten to 200~ps or less \cite{Weinzetl2019,Smallwood2021}. Given that these centers are initially created through silicon-ion bombardment, one may assume that this dramatic reduction in coherence time is caused by disorder of the diamond lattice introduced during SiV$^-$ implantation due to the higher silicon ion flux required to create an ensemble. Recent MDCS experiments on SiV$^-$ ensembles \cite{Smallwood2021} point to a more subtle reality, however, revealing an additional population of non-luminous centers with drastically longer coherence times exceeding 900~ps. Indeed, these two distinct decoherence timescales are difficult to reconcile with decoherence induced by simple lattice disorder, which should impact all centers similarly.


\begin{figure}[b]
    \centering
    \includegraphics[width=0.4\textwidth]{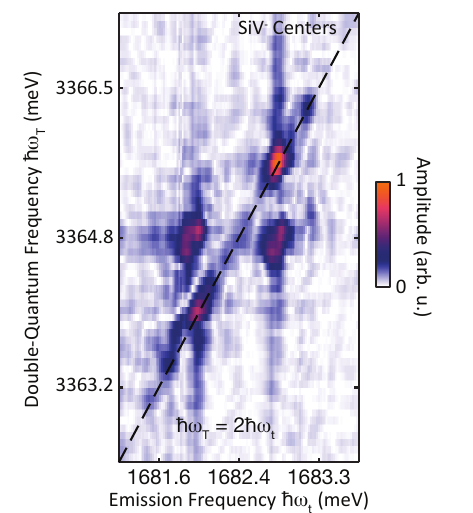}
    \caption{Photoluminescence-detected double-quantum MDCS spectrum acquired at 12 K, demonstrating that pairs of SiV$^-$ centers interact via dipole-dipole interactions \cite{Day2022}. The two peaks along $\hbar\omega_T = 2\hbar\omega_t$ reflect resonant interactions between two SiV$^-$ centers prepared in the same quantum state during measurement while the two remaining peaks correspond to non-resonant interactions between two SiV$^-$ centers prepared in different quantum states.}
    \label{Fig7}
\end{figure}

The underlying mechanism was finally revealed by `double-quantum' MDCS spectra \cite{Day2022}, which crucially exhibit a non-zero signal only in the presence of interactions between members of an ensemble. In the double-quantum spectrum shown in Figure~\ref{Fig7}, the presence of peaks at twice the excitation frequency along the vertical frequency axis $\hbar\omega_T$ thus point to resonantly-enhanced dipole-dipole interactions that lead to interaction-dominated, correlated dephasing of the ensemble as a whole. These interactions, while deleterious to coherence times in the specific sample in question, can potentially be used to control the joint states of SiV$^-$ and color centers more generally. It was further demonstrated that the strength of these interactions can even be varied by modifying the initial state of the ensemble with an additional `pre-pulse' \cite{Day2022}. Tunable interactions can provide a tool for coherent control of the SiV$^-$ ensembles and, from a more general perspective, can be a resource for studying correlation-induced many-body physics in complex physical systems.

\section{Summary and Outlook}

In this Perspective, we have reviewed the concept of decoherence in quantum systems as well as conventional techniques for measuring decoherence. In our view, MDCS overcomes many limitations of conventional techniques and is capable of probing complex decoherence processes central to the development of future quantum technologies. We have demonstrated this with exemplary MDCS studies on vacancy center and colloidal nanocrystals from our own recent research.

We anticipate that advances in two particular directions, (1) integrating MDCS with optical microscopy \cite{Purz2022,Purz2022_2} and (2) extending the frequency range of MDCS into the terahertz frequency range \cite{Lu2019_Chapter}, will greatly accelerate the rational design of material platforms for quantum technologies. Along the former direction, endowing MDCS with spatial resolution will enable correlation of inhomogeneity in an electronic response with inhomogeneity of the underlying material structure. Along the latter direction, MDCS at terahertz frequencies \cite{Liu2025} offers immense potential to probe collective excitations of strongly correlated {\it quantum materials} \cite{Liu_2023_Echo,Salvador2024}, in which decoherence is poorly understood.

\bibliography{bibliography}

\end{document}